\documentclass[12pt]{iopart}

\usepackage{graphicx}
\usepackage{iopams}
\begin{document}

\title{Quantum-to-classical correspondence in open chaotic systems}

\author{Henning Schomerus\dag\ and Philippe Jacquod\ddag}

\address{\dag\ Department of Physics, Lancaster University,
Lancaster LA1 4YB, United Kingdom}
\address{\ddag\ D\'epartement de Physique Th\'eorique,
Universit\'e de Gen\`eve, CH-1211 Gen\`eve 4, Switzerland}

\begin{abstract}
We review properties of open chaotic mesoscopic systems with a
finite Ehrenfest time $\tau_{\rm E}$. The Ehrenfest time separates
a short-time regime of the quantum dynamics, where
 wave packets closely follow
the deterministic classical motion, from a long-time regime of
fully-developed wave chaos. For a vanishing Ehrenfest time the
quantum systems display a degree of universality which is well
described by random-matrix theory. In the semiclassical limit,
$\tau_{\rm E}$ becomes parametrically larger than the scattering
time off the boundaries and the dwell time in the system. This
results in the emergence of an increasing number of deterministic
transport and escape modes, which induce strong deviations from
random-matrix universality. We discuss these deviations for a
variety of physical phenomena, including shot noise, conductance
fluctuations, decay of quasibound states, and the mesoscopic
proximity effect in Andreev billiards.

\end{abstract}


\section{Introduction}
Technological advances of the past decade have made it possible
to construct clean electronic devices of a linear size smaller
than their elastic mean free path, but still much larger than the
Fermi wavelength \cite{Revdot1,Revdot2,Revdot3}. The motion of the
electrons in these {\it quantum dots} is thus ballistic, and
determined by an externally imposed confinement potential. On the
classical level the dynamics can vary between two extremes,
according to whether the confinement potential gives rise to
integrable or chaotic motion \cite{Lichtenberg,Ott}. This
classification is carried over to the quantum dynamics
\cite{Bar93,Mar93,Haake,stoeckmann}, with most of the theoretical
efforts focusing on the chaotic case. It has been conjectured
\cite{BGS,Ber85} that chaotic quantum dots fall into the same
universality class as random disordered quantum dots. The latter
exhibit  a degree of universality which is well captured by
random-matrix theory (RMT) \cite{RMT1,RMT2,Ben97}. Starting, e.g.,
from the scattering theory of transport \cite{scatg1,scatg2}, RMT provides a
statistical description, where the system's scattering matrix is
assumed to be uniformly distributed over one of Dyson's circular
ensembles \cite{blum90,Meh91,Guh98}. This sole assumption allows
to calculate transport quantities such as the average conductance,
shot noise, and counting statistics, including coherent quantum
corrections such as weak localization and universal conductance
fluctuations \cite{RMT1,RMT2,Ben97}. RMT can also be extended to
the energy dependence of the scattering matrix by linking it to a
random effective Hamiltonian of Wigner's Gaussian universality
classes \cite{Ben97,Guh98,efetovbook}. This allows to investigate
quantum dynamical properties such as time delays and escape rates
\cite{FS}, as well as the density of states of
normal-superconducting hybrid structures \cite{Mel96}.

RMT assumes that all cavity modes are well mixed, hence, that the
system displays well-developed wave chaos. Impurities serve well
for this purpose, since s-wave scattering couples into all
directions with the same strength. Indeed, RMT enjoys a well-based
mathematical foundation for disordered systems
\cite{efetovbook,efetov,piet95}. In ballistic chaotic systems, the
scattering is off the smooth boundaries, which is not directly
diffractive in contrast to the scattering from an impurity. On the
other hand, classical chaos is associated to a strong sensitivity
of the dynamics to uncertainties in the initial conditions, which
are amplified exponentially with time $\sim \exp(\lambda t)$
(where $\lambda$ is the Lyapunov exponent). Hence, there are
little doubts that RMT also gives a correct description for
ballistic systems where this sensitivity yields to well-developed
wave chaos.

The precise conditions under which wave chaos emerges
from classical chaos are however just being uncovered. Generically, RMT is
bounded by the existence of finite time scales. In closed chaotic
systems, for instance, spectral fluctuations are known to deviate
from RMT predictions for energies larger than the inverse period
of the shortest periodic orbit \cite{Ber85}. For transport through
open chaotic systems, classical ergodicity clearly has to be
established faster than the lifetime of an electron in the system.
Accordingly, the dynamics should not allow for too many short,
nonergodic classical scattering trajectories going straight
through the cavity, or hitting its boundary only very few times
\cite{Naz03}. This requires that the inverse Lyapunov exponent
$\lambda^{-1}$ and the typical time $\tau_{\rm B}$ between bounces
off the confinement are much smaller than the dwell time
$\tau_{\rm D}$, hence $\lambda^{-1}, \tau_{\rm B} \ll \tau_{\rm
D}$. In practice, these conditions are fulfilled when the openings
are much smaller than the linear system size $L$. RMT universality
also requires that
 $\lambda^{-1}$, $\tau_{\rm B}$, and $\tau_D$ are
smaller than the Heisenberg time $\tau_{\rm H}=h/\Delta$ (with
$\Delta$ the mean level spacing). The condition $\tau_{\rm
H}\gg\tau_{\rm B}$ guarantees that a large number of internal
modes $M={\rm Int}[\tau_{\rm H}/\tau_{\rm B}]$ are mixed by the
chaotic scattering (RMT is then independent of microscopic details
of the ensemble \cite{Ben97}). The condition $\tau_H\gg\tau_D$
translates into a large total number of open scattering channels
in all leads, $N_{\rm tot}={\rm Int}\, [\tau_{\rm H}/\tau_{\rm
D}]$, such that details of the openings can be neglected. Together
with the condition $\tau_B\ll\tau_D$, this implies $1\ll N_{\rm
tot} \ll M$. The limit $M\to\infty$, $M/N_{\rm tot}={\rm const}$
is equivalent to the semiclassical limit of a small Fermi wave
length $\lambda_{\rm F}/L \to 0$, all classical parameters
being kept fixed. These requirements have been thoroughly
investigated in the past \cite{Ben97,piet95,Lew91}.

More recently, following the seminal work of Aleiner and Larkin
\cite{Ale96}, it has become clear that a new time scale,
associated to the {\em quantum-to-classical correspondence} of
wave-packet dynamics (and in this sense, the validity of
Ehrenfest's theorem), also restricts the validity of the RMT of
ballistic transport. This time scale $\tau_{\rm E}$, usually
referred to as the Ehrenfest time, is roughly the time it takes
for the chaotic classical dynamics to stretch an initially narrow
wave packet, of Fermi wavelength $\lambda_{\rm F}$, to some
relevant classical length scale ${\cal L}$ (cf.\ Fig.\
\ref{fig:sketch}). Since the stretching $\propto \exp[\lambda t]$
is exponential in time, one has $\tau_{\rm E}\propto\lambda^{-1}
\ln[{\cal L}/\lambda_{\rm F}]$ \cite{Berman78}. The Ehrenfest time
poses a lower limit to the validity of RMT because wave chaos is
associated to the splitting of wave packets into many partial
waves, which then interfere randomly. In ballistic chaotic
systems, the wave packet splitting is only established when
initial quantum uncertainties blow up to the classical level. For
shorter times, the quantum dynamics still bears the signatures of
classical determinism, which is not captured by RMT.

When $\lambda_{\rm F}$ is decreased, all classical parameters
being kept constant -- the very same semiclassical limit
purportedly required for RMT universality -- $\tau_{\rm E}$
becomes parametrically larger than $\tau_{\rm B}$ and
$\lambda^{-1}$, and indeed may start to compete with the dwell
time $\tau_{\rm D}$. One may thus wonder what is left of the RMT
universality of open systems, and more generally of quantum
effects in that limit. Indeed, there are many instances where
quantum-to-classical correspondence at finite $\tau_{\rm E}$ leads
to strong deviations from the universal RMT behavior. Such
deviations are not only of fundamental interest, but also provide
practical mechanisms to suppress or accentuate quantum properties.
This short review provides a survey of the current knowledge of
the quantum-to-classical correspondence in open ballistic systems,
focusing on the deviations from RMT due to a finite Ehrenfest
time.

We start with a brief general classification of the Ehrenfest time
for different physical situations such as transport, escape, and
closed-system properties (section \ref{ehrenfest}). We then turn
our attention to three specific applications where deviations from
RMT universality occur once the relevant Ehrenfest time is no
longer negligible. First (section \ref{transport}), we discuss
transport properties in a two-terminal geometry.
Quantum-to-classical correspondence is reflected in the
distribution of the transmission eigenvalues, and results in the
suppression of electronic shot noise and the breakdown of
universality for sample-to-sample conductance fluctuations. Second
(section \ref{decay}), we discuss the decay modes (quasi-bound
states) of the system. Escape routes faster than the Ehrenfest
time give rise to highly localized, ballistically decaying
quasi-bound states, while the density of long-lived quasi-bound
states is renormalized according to a fractal Weyl law. Finally
(section \ref{andreev}), we investigate the excitation spectrum of
normal-metallic ballistic quantum dots coupled to an $s$-wave
superconductor (the mesoscopic proximity effect). The presence of
the superconducting terminal introduces a new dynamical process
called {\it Andreev reflection} (charge-inverting
retroreflection), which induces a finite quasiparticle lifetime
and opens up a gap in the density of states around the Fermi
energy. The size and shape of the gap show deviations from the RMT
predictions when the Ehrenfest time is no longer negligible
against the lifetime of the quasiparticle. Conclusions are
presented in section \ref{conclusions}.

Because of the slow, logarithmic increase of $\tau_{\rm
E}\propto\ln M$ with the effective size $M$ of the Hilbert space,
the ergodic semiclassical regime $\tau_{\rm E} \gtrsim \tau_{\rm
D}$, $\lambda \tau_{\rm D} \gg 1$ is unattainable by standard
numerical methods. The numerical results reviewed in this paper
are all obtained for a very efficient model system, the open
kicked rotator \cite{Jac03,Two03,Bor91,Oss02}, which we briefly
describe in the Appendix.

\section{Classification of Ehrenfest times \label{ehrenfest}}

\begin{figure}
\begin{center}
\includegraphics[width=14cm]{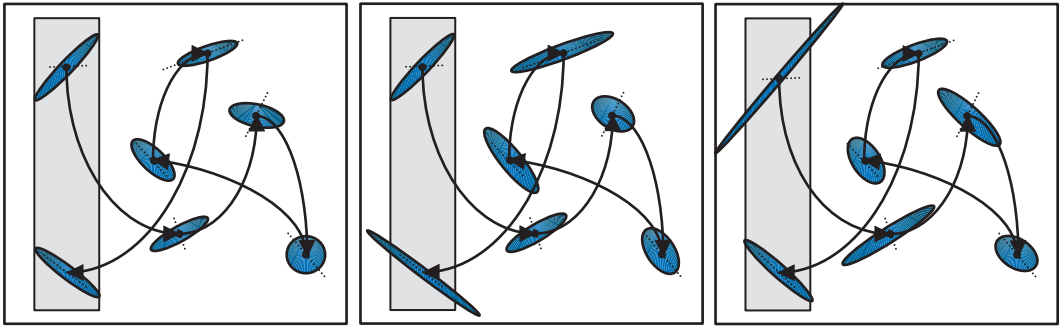}
\end{center}
\caption{\label{fig:sketch} Sketch of the dynamics of a wave
packet in the phase space of an open system. The left and middle
panels apply to the case of transport. The initial wave packet is
maximally stretched along the stable manifold without substantial
leakage out of the shaded rectangular area, which represents an
opening of the system. After five bounces the wave packet returns
to the opening, now being elongated along the unstable manifold
(dashed lines; the sketches neglect the bending of the manifolds).

In the left picture, the transport Ehrenfest time $\tau_{\rm
E}^{(2)}$ is larger than the dwell time $\tau_{\rm D}$. The
returning wave packet still fits through the opening, with only
minimal leakage. Hence, the particle leaves the system
deterministically, as prescribed by the classical dynamics of the
wave-packet center (dots).

The middle picture corresponds to a more chaotic system (with
larger Lyapunov exponent), resulting in $\tau_{\rm E}^{(2)} <
\tau_{\rm D}$. The stretching is stronger and the wave packet is
not fully transmitted. In the subsequent dynamics, the partially
reflected wave components will interfere randomly, which gives
rise to wave chaos.

In the escape problem, the initial wave packet can be squeezed
more closely to the stable manifold, and the associated Ehrenfest
time $\tau_{\rm E}^{(1)}$ is larger than the transport Ehrenfest
time. This is illustrated in the right panel. }
\end{figure}

The relevant Ehrenfest time depends on the physical situation at
hand, but follows a very simple classification.
Quantum-to-classical correspondence is maximized for wave packets
that are initially elongated along the stable manifold of the
classical dynamics, so that the dynamics first yields to
compression, not to stretching (see Fig.\ \ref{fig:sketch}). The
initial extent of elongation along the stable manifolds is limited
either by the linear width of the openings $W$, or the linear
system size $L$, depending on whether the physical process
requires injection into the system or not. In the same way, the
final extent of the wave packet has to be compared to $W$ or $L$
depending on whether the physical process requires the particle to
leave the system or not. For sufficiently ergodic chaotic dynamics
($\lambda^{-1}$, $\tau_{\rm B} \ll \tau_{\rm D}$, which implies
$W\ll L$) and in the absence of sharp geometrical features
(besides the presence of the openings), the resulting Ehrenfest
time can be expressed by the three classical time scales  and the
Heisenberg time $\tau_{\rm H}$ \cite{Berman78,Vav03,Scho04}:
\begin{equation}
\tau_{\rm E}^{(n)}=\lambda^{-1}\ln
\left[\frac{\tau_{\rm H}}{\tau_{\rm B}}\left(\frac{\tau_{\rm B}}{\tau_{\rm D}}\right)^n\right]
\;\;\; ; \;\;\; n=\left\{
\begin{array}{cc}
0 & {\rm closed \; system}, \\
1 & {\rm escape}, \\
2 & {\rm transport}.
\end{array}
\right.
\label{eq:lyapexp}
\end{equation}
Here $n$ gives the number of passages through the openings
associated to the physical process. Expression (\ref{eq:lyapexp})
holds for two-dimensional systems such as lateral quantum dots
\cite{Revdot1,Revdot2,Revdot3} or microwave cavities
\cite{stoeckmann}, as well as for the stroboscopic one-dimensional
model systems often used in the numerical simulations ($\tau_{\rm
B}$ is then the stroboscopic period; see the Appendix). Expression
(\ref{eq:lyapexp}) also holds for three-dimensional systems (such
as metallic grains) with two-dimensional openings when $\lambda$
is replaced by the sum of the two positive Lyapunov exponents.

The difference between the three Ehrenfest times can be attributed
to the additional splitting of a wave packet into partially
transmitted and partially reflected waves at each encounter with
an opening. Transport involves two passages via the openings. The
first passage, at injection, determines the initial spread of the
associated wave packet. The Ehrenfest time is then obtained by
comparing the final spread to the width of the opening at the
second passage, where the electron leaves the system. This results
in the transport Ehrenfest time $\tau_{\rm E}^{(2)}=\lambda^{-1}
\ln[\tau_{\rm H}\tau_{\rm B}/ \tau_{\rm D}^2]$. The same Ehrenfest
time also affects the excitation spectrum of normal-metallic
cavities which are coupled by the openings to an $s$-wave
superconductor, for which the relevant physical process is the
consecutive Andreev reflection of the two quasi-particle species
at the superconducting terminal ($\tau_{\rm E}^{(2)}$ was actually
first derived in that context \cite{Vav03}). In the escape
problem, the electron is no longer required to originate from an
opening. This lifts the restriction on the initial confinement of
the wave packet and allows to squeeze it closer to the stable
manifolds, as its elongation is not limited by the width of the
opening but by the linear system size. Hence, the escape Ehrenfest
time is larger than the transport Ehrenfest time by a factor
$\tau_{\rm D}/\tau_{\rm B}$ in the logarithm: $\tau_{\rm
E}^{(1)}=\lambda^{-1}\ln [\tau_{\rm H} / \tau_{\rm D}]$. This
value is exactly in the middle of the transport Ehrenfest time
$\tau_{\rm E}^{(2)}$ and the conventional Ehrenfest time
$\tau_{\rm E}^{(0)}=\lambda^{-1} \ln[\tau_{\rm H}/\tau_{\rm B}]$
for closed systems \cite{Berman78}, for which initial and final
extents of the wave packet must be compared against the linear
size of the system.

The semiclassical limit is achieved for $\tau_{\rm H}\to\infty$
while the classical time scales $\lambda^{-1}$, $\tau_{\rm B}$,
and $\tau_{\rm D}$ are fixed. The Ehrenfest time then increases
logarithmically with $\tau_{\rm H}/\tau_{\rm B}$, which is
$\propto L/\lambda_{\rm F}$ for two dimensions and $\propto
(L/\lambda_{\rm F})^2$ for three dimensions. In this paper, we
will denote this limit by $M\to\infty$ while keeping $M/N_{\rm
tot}$ fixed, where $M={\rm Int}[\tau_{\rm H}/\tau_{\rm B}]$ is the
effective number of internal modes mixed by the chaotic
scattering, and $N_{\rm tot}={\rm Int}[\tau_{\rm H}/\tau_{\rm
D}]\ll M$ is the total number of open channels.

\section{Quantum-to-classical crossover in transport \label{transport}}

A rough classification distinguishes transport properties whose
magnitude can be expressed by classical parameters  from
quantities that rely on quantum coherence \cite{Ben97}. Examples
of the former class are the conductance $G\sim\rho e v_{\rm F}/V$
(where $\rho$ is the electronic density in an energy interval $eV$
around the Fermi energy, $V$ is the voltage, and $v_{\rm F}$ is
the Fermi velocity) and the electronic shot noise $P\sim P_0=2 e^2
G V$. The latter class is represented by the weak localization
correction to the conductance 
and the universal conductance fluctuations, which in
RMT are both of the order of a conductance quantum $G_0=e^2/h$. In
the presence of a finite Ehrenfest time, a much richer picture
emerges: in particular the sample-to-sample conductance
fluctuations are elevated to a classical level, while the shot
noise is suppressed.

The origin of these strong deviations from universal RMT behavior
can be traced down to the distribution of transmission
eigenvalues. We specifically consider transport through a chaotic
cavity in a two-terminal geometry, and restrict ourselves to the
case where the number of open channels leading to the electronic
reservoirs are the same, $N_{\rm L} = N_{\rm R} \equiv N = N_{\rm
tot}/2$. The scattering matrix ${\cal S}$ is a $2N \times 2N$
matrix, written in terms of $N \times N$ transmission (${\bf t}$
and ${\bf t}'$) and reflection (${\bf r}$ and ${\bf r}'$) matrices
as
\begin{eqnarray}\label{blocks}
{\cal S}= \left( \begin{array}{ll}
{\bf r} & {\bf t}' \\
{\bf t} & {\bf r}'
\end{array}\right).
\end{eqnarray}
The system's conductance is given by $G/G_0 = {\rm Tr} ({\bf
t}^\dagger {\bf t}) = \sum_n T_n$ \cite{scatg1,scatg2}, where the
$T_n\in [0,1]$ are the transmission eigenvalues of ${\bf
t}^\dagger {\bf t}$. In the limit $N \rightarrow \infty$ and
within RMT, their probability distribution is given by
\cite{RMT1,RMT2,Ben97}
\begin{equation}\label{probt}
P_{\rm RMT}(T) = \frac{1}{\pi} \frac{1}{\sqrt{T(1-T)}} \;\;\;\;\;\;
; \;\;\;\;\;\; T \in[0,1].
\end{equation}

\begin{figure}
\begin{center}
\includegraphics[width=14cm]{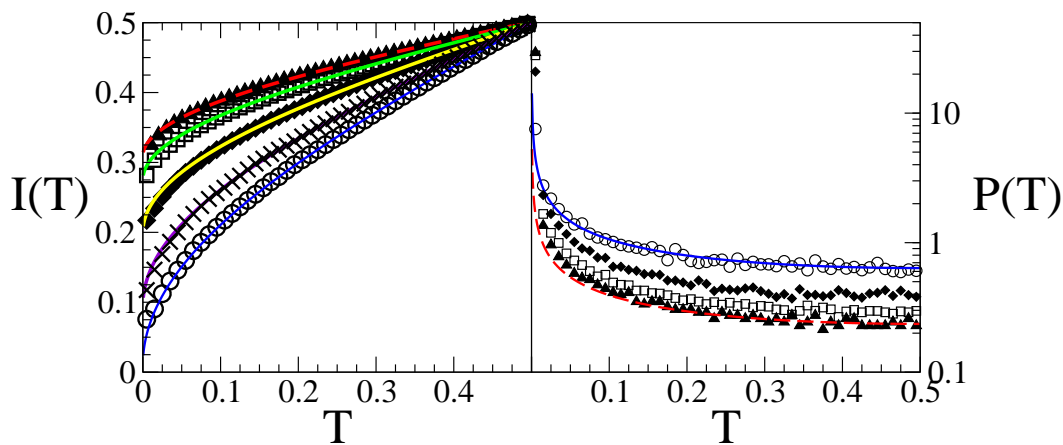}
\end{center}
\caption{\label{transmission} Left panel: Integrated probability
distribution $I(T)$ of transmission eigenvalues for open kicked
rotators with $\tau_{\rm D}/\tau_{\rm B}=25$ and $\tau_{\rm
E}^{(2)} \simeq 0$ (empty circles; distribution calculated over
729 different samples); $\tau_{\rm D}/\tau_{\rm B}=5$, and
$\tau_{\rm E}^{(2)}/\tau_{\rm B}=0.16$ ($\times$; 1681 samples),
$\tau_{\rm E}^{(2)}/\tau_{\rm B}=1.5$ (black diamonds;  729
samples), $\tau_{\rm E}^{(2)}/\tau_{\rm B}=2.8$ (empty squares; 16
samples), and $\tau_{\rm E}^{(2)}/\tau_{\rm B}=4.1$ (black
triangles; 2 samples). The blue, violet, yellow, green and red
lines give the distribution $I_{\alpha}$ of
Eq.~(\ref{iprobtalpha}), with $\alpha \approx 0.98$, 0.81, 0.6,
0.45 and 0.385 respectively. Right panel: Probability distribution
$P(T)$ of transmission eigenvalues for the same set of parameters
as in the main panel (data for $\tau_{\rm D}/\tau_{\rm B}=5$ and
$\tau_{\rm E}^{(2)}/\tau_{\rm B}=1.5$ have been removed for
clarity). The blue solid line gives the universal distribution
$P_{\rm RMT}$ of Eq.~(\ref{probt}) while the red dashed line
corresponds to Eq.~(\ref{probtalpha}), with $\alpha = 0.39$. Note
that $P(T)$ is symmetric around $T=0.5$.}
\end{figure}

Equation (\ref{probt}) requires that the standard conditions for RMT universality discussed in the
introduction are met, $\lambda^{-1},\tau_{\rm B}\ll\tau_{\rm D}\ll\tau_{\rm H}$ (hence, $1\ll N \ll M$).
However, even when those conditions apply, it has recently
been observed that
strong deviations from Eq.~(\ref{probt}) occur
in the semiclassical limit $M\to\infty$ \cite{Jac04}.
This is illustrated in Fig.~\ref{transmission}, which shows the results of a numerical investigation of
the open kicked rotator (described in the Appendix).
Instead of Eq.~(\ref{probt}),
the transmission eigenvalues appear to be distributed according to
\begin{equation}\label{probtalpha}
P_{\rm \alpha}(T) = \alpha P_{\rm RMT}(T) +\frac{1-\alpha}{2}
\left[\delta(T)+\delta(1-T)\right].
\end{equation}
The presence of $\delta$-peaks at $T=0$ and $T=1$ in
$P_{\rm \alpha}(T)$ becomes more evident once the integrated distribution
$I(T)=\int_0^T P(T') dT'$ is plotted. From Eq.~(\ref{probtalpha}) one has
\begin{equation}\label{iprobtalpha}
I_{\rm \alpha}(T) = \frac{2 \alpha}{\pi} \arcsin \sqrt{T} \; + \;
\frac{1-\alpha}{2} (1 + \delta_{1,T}),
\end{equation}
so that $I_{\rm \alpha}(0)=(1-\alpha)/2$ vanishes only for
$\alpha=1$. For the data in Fig.~\ref{transmission}, it turns out
that the parameter $\alpha$ is well approximated by $\alpha
\approx \exp[-(\tau_{\rm E}^{(2)}+\tau_{\rm B})]/\tau_{\rm D}]$
\cite{Jac04}, with the transport Ehrenfest time $\tau_{\rm
E}^{(2)}$ given in Eq.\ (\ref{eq:lyapexp}). Hence, for a
classically fixed configuration (i.e. considering an ensemble of
systems with fixed $\lambda$, $\tau_{\rm B}$, and $\tau_{\rm D}$),
the fraction $f=1-\alpha$ of deterministic transmission
eigenvalues with $T=0,1$ approaches $f=1$ in the semiclassical limit
$M\to\infty$, $M/N={\rm const}$.

The emergence of classical determinism in the semiclassical limit
reflects the fact that short trajectories are able to carry a wave
packet in one piece through the system, provided that the wave
packet is localized over a sufficiently small volume (see Fig.\
\ref{fig:sketch}). Equation (\ref{probtalpha}) moreover suggests
that the spectrum of transmission eigenvalues is the sum of two
independent contributions, precisely what would happen if the
total electronic fluid of the system would split into two
coexisting phases, a classical and a quantal one \cite{caveat2}.
This splitting leads to a two-phase fluid model. It has been
surmised that the quantal phase can be modelled by RMT
\cite{Sil03}, which results in an {\it effective random-matrix
model} with renormalized matrix dimension $\alpha N$. Since
$\alpha N\propto N^{1-1/\lambda\tau_{\rm D}}\to\infty$ in the
semiclassical limit (see the related discussion of the fractal
Weyl law in section \ref{decay}), effective RMT predicts that the
universality of quantum interference such as weak localization and
parametric conductance fluctuations is not affected by a finite
Ehrenfest time. This model is supported by a semiclassical theory
based on the two-fluid model \cite{Whi05}. On the other hand, a
stochastic quasiclassical theory which models mode-mixing by isotropic
residual diffraction predicts that quantum interference effects
are suppressed for a finite Ehrenfest time
\cite{Ale96,Bro05,Bro05b,Ada03}.

Numerical investigations on parametric conductance 
fluctuations \cite{Jac04,Bro05b,Two04cond} give support for the RMT universality
of the quantal phase  (see Sec.\ \ref{sec:ucf}),
and variants of the effective RMT 
model have been successfully utilized beyond transport 
applications (see Secs.\ \ref{decay}, \ref{andreev}).
On the other hand, while an
earlier numerical investigation of the weak-localization 
correction \cite{Two04} reported no clear dependence of the 
magnetoconductance $\delta G=G(B=0)-G(B=\infty)$
on the Ehrenfest time, very recent investigations \cite{Bro05,Bro05b} find 
a suppression of $\delta
G$ for an increasing Ehrenfest time. The observed suppression is in agreement with the
prediction $\delta G \propto \exp[-\tau_{\rm E}^{(2)}/ \tau_{\rm
D}]$ of a modified quasiclassical theory \cite{Bro05b} in which
the suppression results from electrons 
with dwell time between $\tau_{\rm E}^{(2)}$ and $2\tau_{\rm E}^{(2)}$.
However, the quasiclassical theory cannot explain why the parametric conductance 
fluctuations are not suppressed. It also does not yet deliver as many predictions beyond
transport as the effective RMT (for quasiclassical predictions of the mesoscopic 
proximity effect, see Sec.\ \ref{andreev}).

At present, both effective RMT as well as the quasiclassical theory
have to be considered as phenomenological models, as they involve
uncontrolled approximations.
Clearly, a microscopic theory for the quantal phase which
establishes the extent of its universality is highly desirable.
This poses a considerable theoretical challenge considering that
even in the limit of a vanishing Ehrenfest time a microscopic
foundation for RMT in ballistic systems is only slowly emerging
\cite{Richter}. In this section, we focus
on the consequences of the emergence of deterministic transport
modes for the shot noise and the conductance fluctuations, as most
of these consequences are largely independent of the precise
degree of universality among the non-deterministic transport
modes.

\subsection{Suppression of shot noise}
Shot noise is the non-thermal component of the electronic current fluctuations
\cite{Bla00}
\begin{equation}
P(\omega)=2\int_{-\infty}^\infty \langle \delta I(t)\delta I(0) \rangle \exp(i\omega t)\,dt
,
\end{equation}
where $\delta I(t) = I(t)-\bar I$ is the deviation of the current
from the mean current $\bar I$, and $\langle \ldots \rangle$
denotes the expectation value. This noise arises because of
stochastic uncertainties in the charge carrier dynamics, which can
be caused by a random injection process, or may develop during the
transport through the system. For completely uncorrelated charge
carriers, the noise attains the Poissonian value $P_0=2 e^2 G V$.
Deviations from this value are a valuable indicator of
correlations between the charge carriers.

Phase coherence requires sufficiently low temperatures, at which
Pauli blocking results in a regular injection and collection of
the charge carriers from the bulk electrodes. The only source of
shot noise is then the quantum uncertainty with which an electron
is transmitted or reflected. This is expressed by the quantum
probabilities $0 \le T_n \le 1$. In terms of these probabilities,
the zero-frequency component of the shot noise is given by
\cite{buettiker}
\begin{equation}
P(\omega=0)=2 G_0 e V\sum T_n(1-T_n).
\label{eq:noise}
\end{equation}
This is always smaller than the Poisson value,which can be attributed to the Pauli blocking.

In RMT, the universal value of shot noise in cavities with
symmetric openings follows from Eq.~(\ref{probt}),
$P(\omega=0)=\frac{1}{4}P_0$ \cite{RMT1}. It was predicted by Agam
{\em et al.} \cite{Aga00} that shot noise is further reduced below
this value when the Ehrenfest time is finite,
\begin{equation}
 P(\omega=0)=\frac{1}{4}P_0\exp(-\tau_{\rm E}/\tau_{\rm D}).
\end{equation}
The  RMT value has been observed by Oberholzer {\em et al.} in
shot-noise measurements on lateral quantum dots \cite{Obe00}. The
same group later observed that the shot noise is reduced below the
universal RMT result when the system is opened up (which reduces
$\tau_{\rm D}/\tau_{\rm B}$, not $\tau_{\rm H}/\tau_{\rm B}$)
\cite{Obe02,BeenToday}.

\begin{figure}
\begin{center}
\includegraphics[width=10.5cm]{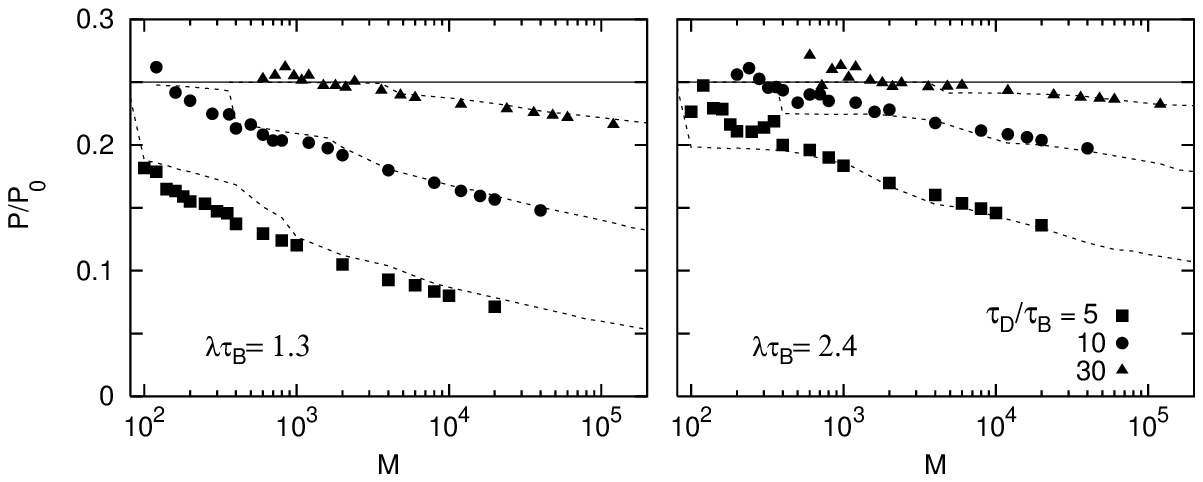}
\includegraphics[width=5cm]{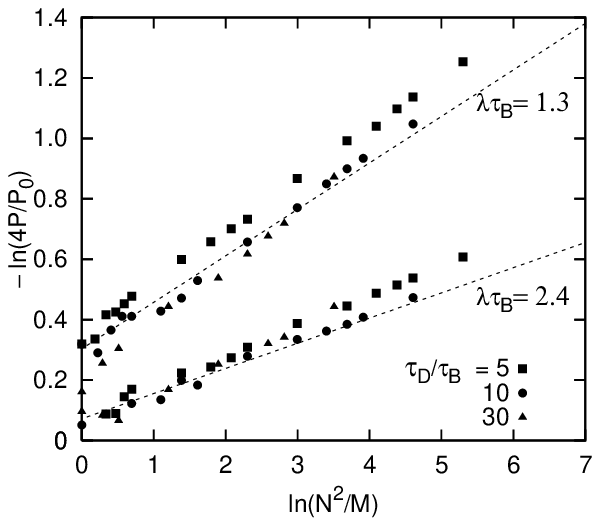}
\end{center}
\caption{\label{fig:noise} Left and middle panels: dependence of
the shot noise on the semiclassical parameter $M\sim \tau_{\rm
H}/\tau_{\rm B}$ for two different Lyapunov exponents $\lambda$
and three different dwell times $\tau_{\rm D}$ of the open kicked
rotator (as indicated in the plots). The solid line is the
prediction from RMT. The dashed lines are obtained from  a
semiclassical estimate of the number of deterministic transport
channels \cite{Sil03}. Right panel: rescaled data in a
double-logarithmic plot, together with lines of slope
$1/\lambda\tau_{\rm D}$. Figures adapted from Ref.~\cite{Two03}.}
\end{figure}

Equation (\ref{eq:noise}) certifies that classically deterministic
transport channels with $T_n=0$ or $T_n=1$ do not contribute to
the shot noise \cite{Bee91}. Ref.\ \cite{Aga00} is based on the
quasiclassical theory \cite{Ale96} which models mode-mixing by
residual diffraction, and equates the Ehrenfest time with the
closed-system Ehrenfest time $\tau_{\rm E}^{(0)}$. The discussion
of the formation of the deterministic transport channels suggests
that this has to be replaced by the transport Ehrenfest time
$\tau_{\rm E}^{(2)}$ \cite{Vav03}. Subsequent numerical
investigations have tested this prediction  for the open kicked
rotator \cite{Two03}. Results for various degrees of chaoticity
(quantified by the Lyapunov exponent $\lambda$) and dwell times
$\tau_{\rm D}$ are shown in Fig.\ \ref{fig:noise}. The shot noise
is clearly suppressed below the RMT value as the semiclassical
parameter $M$ is increased. The right panel shows a plot of $-\ln
(4P/P_0)$ as a function of $\ln N^2/M\sim \ln (\tau_{\rm
H}\tau_{\rm B}/4\tau_{\rm D}^2)$. The data is aligned along lines
with slope $1/\lambda\tau_{\rm D}$. This confirms that the
suppression of the shot noise is governed by $\tau_{\rm E}^{(2)}$,
in agreement with the distribution (\ref{probtalpha}) of
transmission eigenvalues.

\subsection{From universal to quasiclassical conductance fluctuations \label{sec:ucf}}

Universal conductance fluctuations are arguably one of the most
spectacular manifestations of quantum coherence in mesoscopic
systems \cite{ucf}. In metallic samples, the universality of the
conductance fluctuations manifests itself in their magnitude,
${\rm rms}\, G = O(G_0)$, independently on the sample's shape and
size, its average conductance or the exact configuration of the
underlying impurity disorder. In ballistic chaotic systems, a
similar behavior is observed, which is captured by RMT
\cite{RMT1,RMT2,Ben97}. At the core of the universality lies the
{\it ergodic hypothesis} that sample-to-sample fluctuations are
equivalent to fluctuations induced by parametric variations (e.g.
changing the Fermi energy or applying a magnetic field) within a
given sample \cite{ucf}.

\begin{figure}
\vspace{2.cm}
\begin{center}
\includegraphics[width=9.5cm]{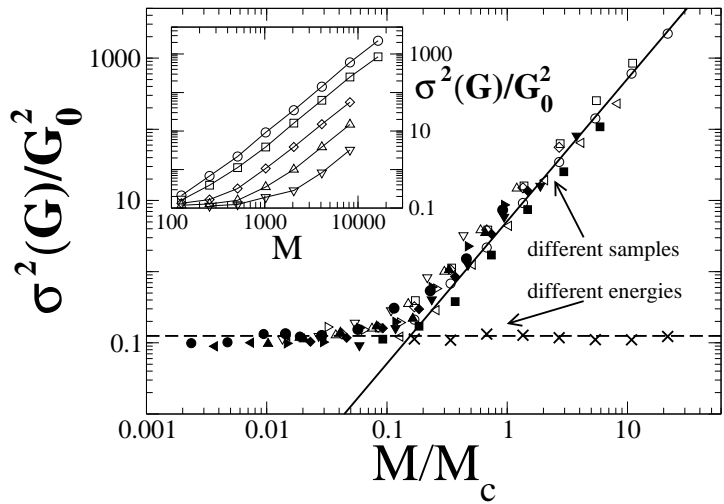}
\end{center}
\caption{\label{fig:ucf} Variance $\sigma^2(G)$ of the conductance
vs. the rescaled effective Hilbert space size $M/M_c$ in the open
kicked rotator for parameters $K \in [9.65,27.65]$, $\tau_{\rm
D}/\tau_{\rm B} \in [5,25]$, and $M \in [128,16384]$. The scaling
parameter $M_c= 2\pi(\tau_{\rm D}/\tau_{\rm B})^2
\exp(\lambda\tau_{\rm B})$ varies by a factor $70$. The solid and
dashed lines indicate the classical, sample-to-sample behavior
$\propto M^2$, and the universal behavior  $\sigma^2(G)=G_0^2/8$,
respectively. Inset: unscaled data for $K=9.65$ and $\tau_{\rm D}/
\tau_{\rm B}=5$ (circles), 7 (squares), 10 (diamonds), 15 (upward
triangles) and 25 (downward triangles). Figure taken from
Ref.~\cite{Jac04}.}
\end{figure}

Three numerical works explored the quantum-to-classical crossover
of conductance fluctuations \cite{Jac04,Bro05b,Two04cond}. Their
findings are consistent with each other and support the conclusion
that (i) the ergodic hypothesis breaks down once $\tau_{\rm
E}^{(2)}$ is no longer negligible; (ii) under variation of a
quantum parameter such as the energy, the conductance fluctuations
stay at their universal value, independently on $\tau_{\rm
E}^{(2)}/\tau_{\rm D}$; and (iii) sample-to-sample fluctuations
increase sharply above the universal value, $\sigma^2(G) \propto
G_0^2(M/M_c)^2$, when $M$ becomes larger than $M_c =
2\pi(\tau_{\rm D}/\tau_{\rm B})^2 \exp(\lambda\tau_{\rm B})$.

Findings (i)-(iii) are illustrated in Fig.~\ref{fig:ucf}, which presents results obtained from the open kicked rotator model.
 They can
be understood on the basis of the two-phase dynamical fluid discussed above.
The deterministic transport channels are insensitive to the variation
of a quantum parameter that influences the phase accumulated on
an unchanged classical trajectories. However, once one changes the
sample configuration, all classical trajectories are scrambled and
huge classical conductance fluctuations result.
The size of the classical fluctuations is determined by the
quantum mechanical resolution of classical phase space structures
corresponding to the largest cluster of fully transmitted or reflected
neighboring trajectories (see Ref.~\cite{Sil03}), which yields the scaling with $M$
 (inset of Fig.~\ref{fig:ucf}) and $M_c$ (main panel of Fig.~\ref{fig:ucf}).
When a quantum parameter is varied, the
conductance fluctuates only due to
long, diffracting trajectories with $t>\tau_{\rm E}^{(2)}$, which build
 up the quantal phase. With the further
assumption that the quantal phase is described by the effective
RMT model, it follows that the  parametric fluctuations are
universal, independently on $\tau_{\rm E}^{(2)}$ (crosses in the
main panel of Fig.~\ref{fig:ucf}). These conclusions are also
supported by the observation in  Ref.~\cite{Jac04} that the energy
conductance correlator $F(\varepsilon)  = \sigma^{-2} (G) \;
\langle \delta G(\varepsilon_0) \delta
G(\varepsilon_0+\varepsilon) \rangle$  decays on the universal
scale of the Thouless energy, $\propto h/\tau_{\rm D}$,
independently on $\tau_{\rm E}^{(2)}$.

\section{Decay of quasi-bound states \label{decay}}
Suppose that the particle is not injected by one of the openings
but is instead prepared (e.g., as an excitation) inside the system
and then escapes through the openings (we will consider the case
of a single opening with $N_{\rm tot}\equiv N$ channels). Instead
of the transport modes, this situation leads us to consider the
decay modes of the system, determined by the stationary
Schr{\"o}dinger wave equation with outgoing boundary conditions.
In contrast to the hermitian eigenvalue problem for a closed
system, an open system with such boundary conditions features a
non-selfadjoint Hamilton operator with complex energy eigenvalues
and mutually non-orthogonal eigenmodes, called quasi-bound states
\cite{Guh98,FS,narimanov}. The imaginary part of the complex energy
$E=E'-i\frac{\hbar \Gamma}{2}$ of a quasibound state is associated
to its escape rate $\Gamma$  (hence, all eigenvalues lie in the
lower half of the complex-energy plane). These energies coincide
with the poles of the scattering matrix, which establishes a
formal link between transport and escape. Since RMT encompasses
also the energy dependence of the scattering matrix, it delivers
precise predictions for the escape rates and wave functions of the
quasi-bound states. Hence, we are again confronted with the issue
to determine the range of applicability of these predictions in
light of the signatures of classical determinism observed in the
short-time dynamics up to the characteristic Ehrenfest time for
the escape problem.

The universal RMT prediction for the escape rates can be obtained via two routes.
The standard route relates
the scattering matrix ${\cal S}$ to an effective
$M \times M$-dimensional Hamiltonian matrix $H$ representing the closed billiard, and
$M \times N$-dimensional matrices $W$ that couple it to the openings \cite{Guh98},
\begin{equation}\label{heidelberg}
{\cal S}(E) = 1 - 2 \pi i W^T (E-H+i \pi W W^T)^{-1} W.
\end{equation}
The superscript ``$T$'' indicates the transpose of the matrix. The
poles of the scattering matrix are then obtained as the
eigenvalues of the non-hermitian matrix $H-i \pi W W^T$. Assuming
that $H$ is a random Gaussian matrix, one can obtain detailed
predictions of the density of these eigenvalues for arbitrary
coupling strength \cite{FS}. For $1\ll N\ll M$, the probability
density of decay rates is then given by
\begin{equation}
P(\Gamma)=\frac{1}{\tau_{\rm D}\Gamma^2}\Theta(\Gamma-\tau_{\rm D}^{-1}),
\label{eq:pgamma}
\end{equation}
where $\Theta$ is the unit step function.

The second route to the distribution of decay rates is particularly adaptable for the case of
ballistic dynamics. It starts from the formulation of the scattering matrix in
terms of
an {\em internal} $M \times M$-dimensional scattering matrix $U(E)$,
which describes the return amplitude to the confinement of the system
\cite{prange,almeida,fyodorov}. For ballistic openings one has
\begin{equation}\label{smatrixdecay}
{\cal S}(E) = P U(E) [1 - (1-P^T P)U(E) ]^{-1} P^T,
\end{equation}
where $P\propto W$
such that $P^TP$ is an idempotent projector onto the leads.
The poles of the scattering matrix are obtained as the
solutions of the determinantal equation ${\rm det}\, [1-(1-P^T P)U(E) ]=0$.

In the semiclassical limit $M\to\infty$, the matrix $U(E)$ carries
an overall phase factor $\exp(i\nu(E))$ with phase velocity
$d\nu/dE=\tau_{\rm B}/\hbar$, equivalent to a level spacing
$\Delta=h/M\tau_{\rm B}=h/\tau_{\rm H}$. In RMT, the distribution
of the poles is obtained under the assumption that $U(E)=F\exp(i E
\tau_{\rm B}/\hbar)$ is proportional to a random unitary matrix
$F$. The truncated unitary matrix $(1-P^T P)F$ has eigenvalues
$\mu=\exp[-(i E'/\hbar+\gamma/2)\tau_{\rm B}]$, where the decay
constant $\gamma$ is related to the decay rate by
$\Gamma=\frac{1}{\tau_{\rm B}}\left(1-e^{-\gamma\tau_{\rm
B}}\right)$. The distribution of these decay constants is given by
\cite{zyc}
\begin{equation}
P(\gamma)=\frac{\tau_{\rm B}^2}{4\tau_{\rm
D}\sinh^2(\gamma\tau_{\rm B}/2)} \Theta(1-e^{-\gamma\tau_{\rm
B}}-\tau_{\rm B}/\tau_{\rm D}),
\end{equation}
which is equivalent to Eq.\ (\ref{eq:pgamma}).

\begin{figure}
\begin{center}
\includegraphics[width=3.5cm]{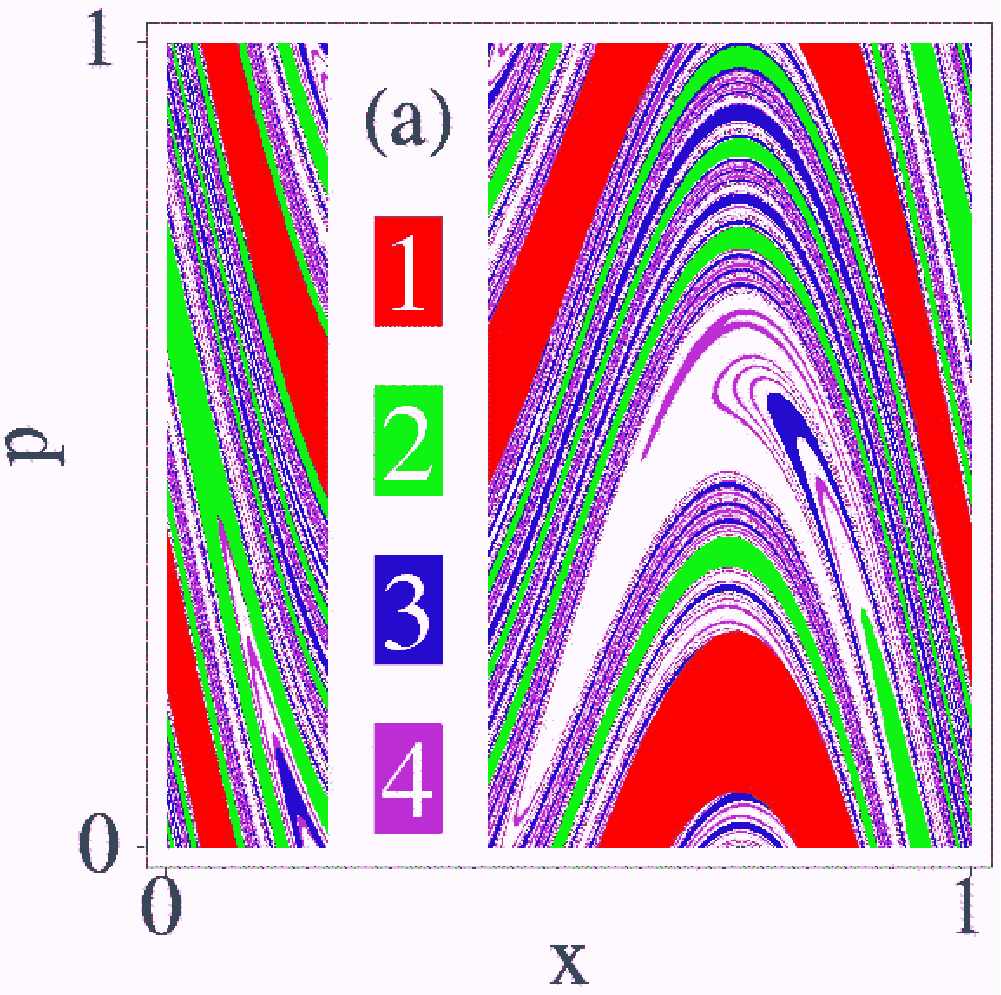}
\includegraphics[width=3.5cm]{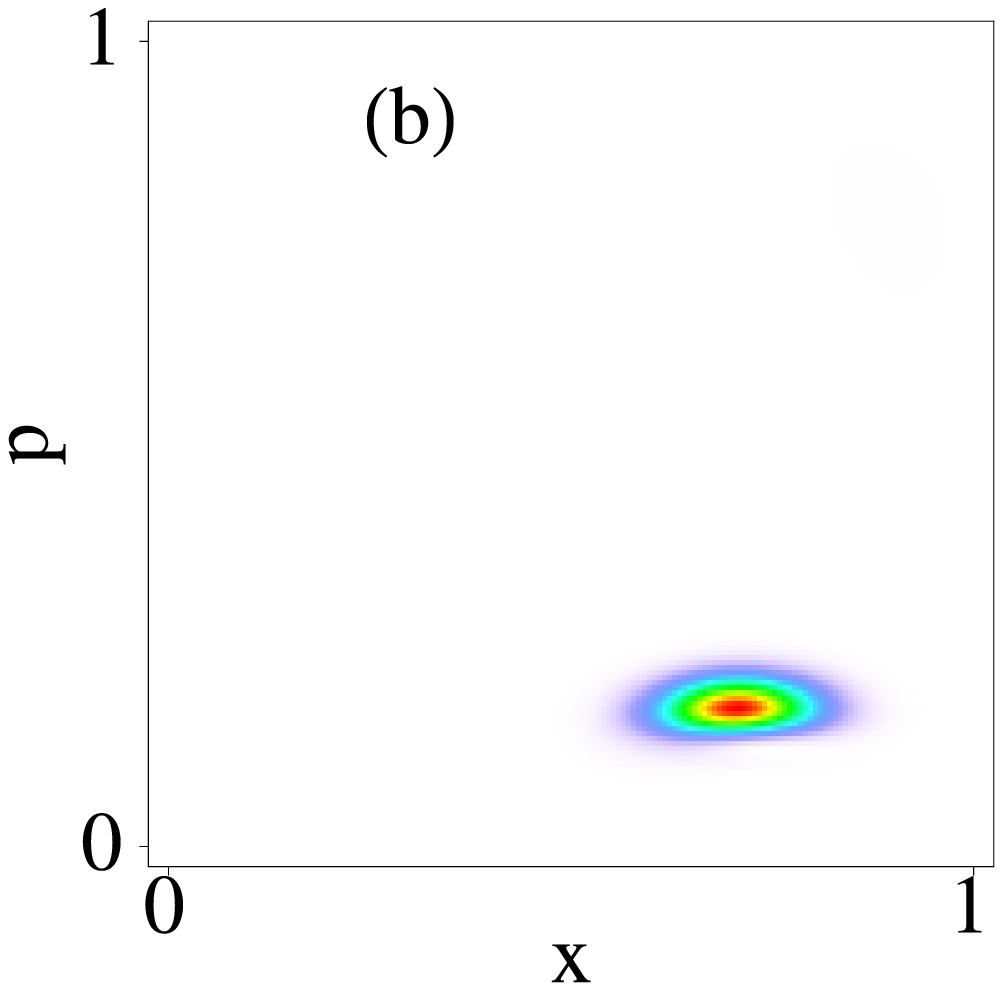}
\includegraphics[width=3.5cm]{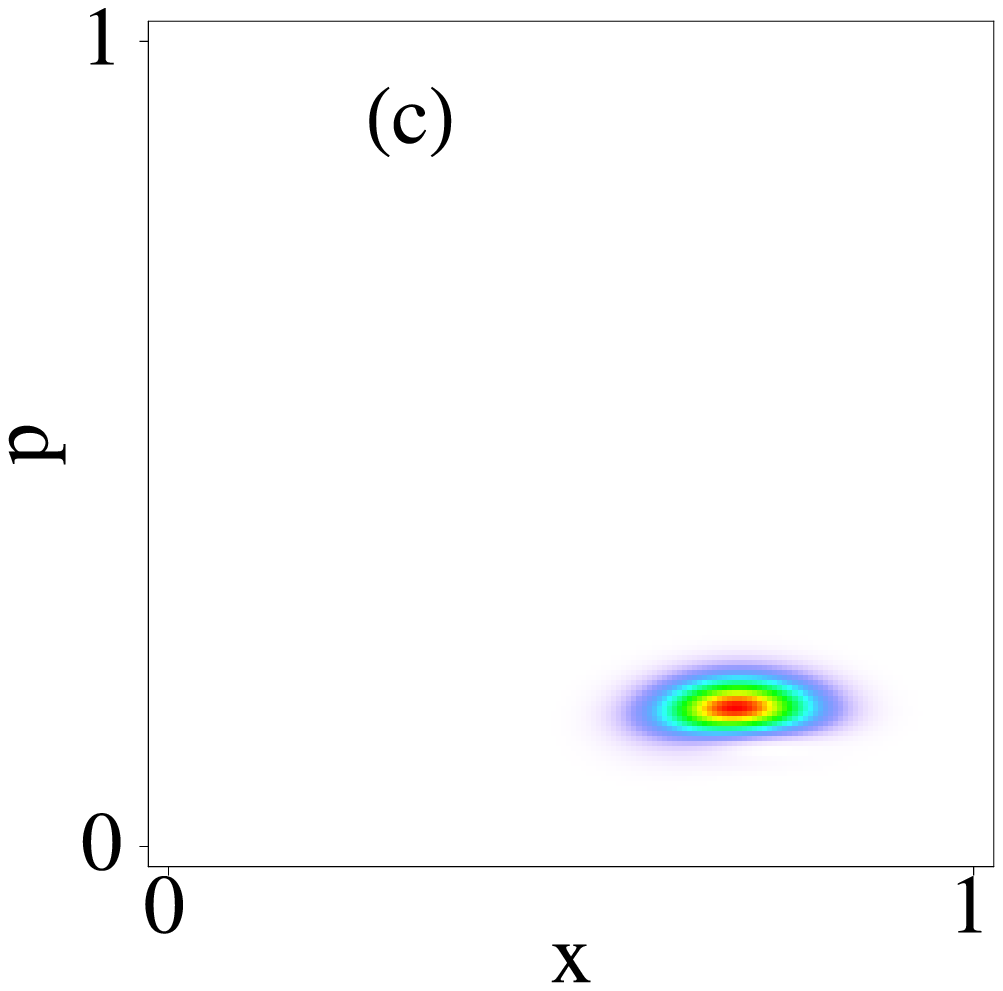}
\includegraphics[width=3.5cm]{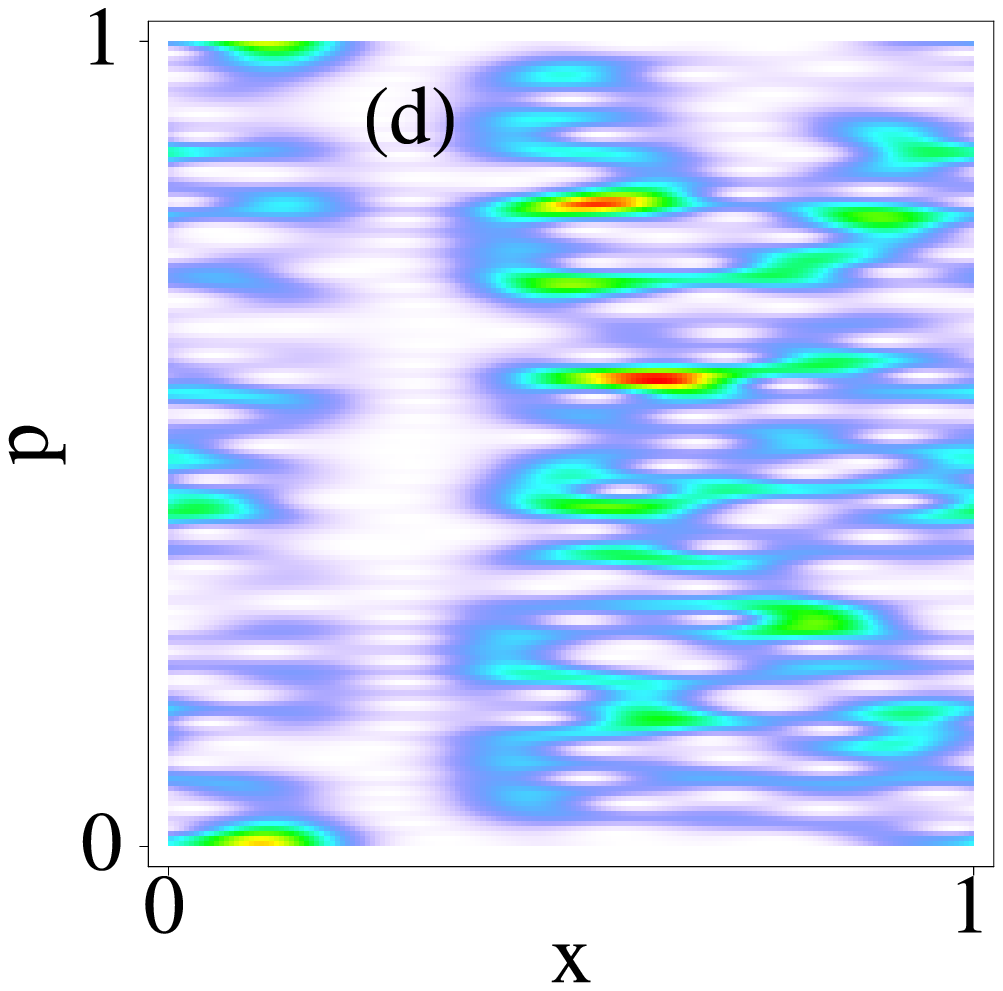}
\end{center}
\caption{\label{fig:decay1} Panel (a): Classical regions of escape
after one to four bounces (color code) in an open kicked rotator
with $\tau_{\rm D}/\tau_{\rm B}=5$ and $K=7.5$ ($\lambda\tau_{\rm
B}\approx 1.3$). Panels (b) and (c): Husimi representations of two
anomalously decaying quasi-bound states for $M=160$. Panel (d):
Husimi representations of a slowly-decaying quasi-bound state.
Figure adapted from Ref.\ \cite{Scho04}.}
\end{figure}

RMT does not account for escape routes with a lifetime shorter
than the Ehrenfest time. Recently, it has been found that these
routes induce the formation of anomalously decaying quasi-bound
states, with very large escape rate $\Gamma$ \cite{Scho04}. The
semiclassical support of the associated wave functions is
concentrated in a small area of phase space, (their total number
is much larger than according to Weyl's rule of one state per
Planck cell). For illustration, Fig.~\ref{fig:decay1} shows
classical regions of escape after a few bounces in the open kicked
rotator, along with two examples of anomalously decaying
quasi-bound states, which are both localized in the same region of
classical escape after one bounce. These states are contrasted
with a slowly decaying state, which displays a random wave
pattern.

In order to discuss these observations, let us assume that the
particle is initially represented by a localized wave packet
$\chi_0$ (such as sketched in  Fig.\ \ref{fig:sketch}). The
evolution of this wave packet from bounce to bounce with the
confinement is given by
\begin{equation}
\chi_m=[(1-P^T P)U(E)]^m\chi_0.
\label{eq:scprop}
\end{equation}
When the final wave packet fits well through the opening the decay is sudden, hence not exponential at all.
For such sudden escape the wave packets generated by the dynamical evolution all are associated to rather special eigenstates of the truncated operator $(1-P^T P)U(E)$: If the escape occurs after $n$ bounces, then
\begin{equation}
[(1-P^T P)U(E)]^{n-m}\chi_m = 0
\label{eq:scleak}
\end{equation}
(neglecting the exponentially suppressed leakage out of the opening area). This corresponds to a highly degenerate eigenvalue $\mu=0$,
hence, $\Gamma=\infty$.

Obviously, the (algebraic) multiplicity  of this eigenvalue is at
least $m$. However, in the semiclassical construction  there is
only one true eigenstate associated to this eigenvalue, namely,
$\chi_{n-1}$. This deficiency is a consequence of the
non-normality of the truncated unitary operator, for which the
existence of a complete set of eigenvectors is not guaranteed. The
degeneracy of the states is lifted beyond the semiclassical
approximation, due to leakage out of the opening area. Hence, in
practice one finds a complete set of eigenstates associated to
this escape route, but the states are almost identical and hence
are all supported by the same area in phase space.

As a consequence of the strong overlap of the anomalously decaying
states, Weyl's rule of covering the support of the states by
Planck cells (of size $\sim 1/M$) cannot be used to estimate their
number. The $m$ states $\chi_{n}$, however, are semiclassically
orthogonal, and Eqs.\ (\ref{eq:scprop})  and (\ref{eq:scleak})
imply that they span the same eigenspace as the nearly degenerate
eigenstates (they provide a Schur decomposition). The
orthogonality of these states reinstates the applicability of
Weyl's rule. When we further observe that the semiclassical
construction requires a reliable quantum-to-classical
correspondence of the wave packet dynamics ($m\tau_{\rm
B}<\tau_{\rm E}^{(1)}$), one can estimate the relative fraction
$f$ of ballistically decaying states by the probability to escape
faster than $\tau_{\rm E}^{(1)}$. Under the assumption of  well
developed classical ergodicity ($\tau_{\rm D}\gg\tau_{\rm
B},\lambda^{-1}$), this probability is given by
\begin{equation}
f=1-\exp(-\tau^{(1)}_{\rm E}/\tau_{\rm D}), \label{eq:f}
\end{equation}
with the escape Ehrenfest time given in Eq.\ (\ref{eq:lyapexp})
\cite{Scho04}.

\begin{figure}
\begin{center}
\includegraphics[width=12cm]{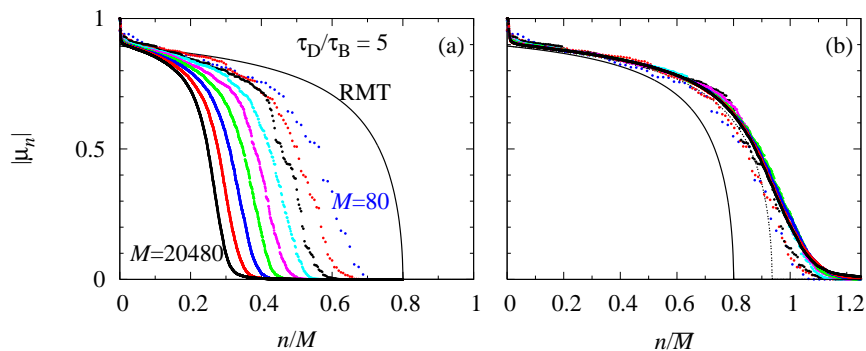}
\end{center}
\caption{\label{fig:decay2} Ordered decay factors
$|\mu_n|=\exp(-\gamma_n\tau_B/2)$ for an open kicked rotator with
$\tau_{\rm D}/\tau_{\rm B}=5$ and $K=7.5$ ($\lambda\tau_{\rm
B}\approx 1.3$) as a function of the relative indices $n/M$ [panel
(a)] and $n/\overline M$ [panel (b)], for $M=2^m\times 80$,
$m=0,1,2,\ldots,8$. The solid line in panel (a) is the RMT result
of Eq.\ (\ref{eq:zycz}) with $\overline{M}=M$. The solid line in
panel (b) is the same result with $\overline M$ given by
Eq.~(\ref{eq:Mbar}--\ref{eq:weyl}). For the dashed line, this
effective dimension has been fitted to the data. Figure adapted
from Ref.\ \cite{Scho04}.}
\end{figure}

Equation (\ref{eq:f}) has been tested numerically in the open
kicked rotator by sorting all decay factors
$|\mu_n|=\exp(-\gamma_n\tau_{\rm B}/2)$, $n=1,2,3,\ldots,M$,
according to their magnitude (see Fig.\ \ref{fig:decay2}). The
data approximately collapses onto a single curve when the relative
index $n/M$ is rescaled with $\exp(-\tau^{(1)}_{\rm E}/\tau_{\rm
D})$. For small decay rates, the scaling function follows closely
the RMT curve \cite{zyc}
\begin{equation}
n(|\mu|)=\overline M[1-\frac{\tau_{\rm B}}{\tau_{\rm
D}}(1-|\mu|^2)^{-1}] \quad (|\mu|^2>1-\tau_{\rm B}/\tau_{\rm D}),
\label{eq:zycz}
\end{equation}
where the matrix dimension is rescaled to
\begin{equation}\label{eq:Mbar}
\overline M=M\exp(-\tau^{(1)}_{\rm E}/\tau_{\rm D}) .
\end{equation}
This equation can be rewritten as
\begin{equation}
\overline M=M^{1-1/\lambda\tau_{\rm D}} (\tau_{\rm D}/\tau_{\rm B})^{1/\lambda\tau_{\rm D}},
\label{eq:weyl}
\end{equation}
which is precisely of the form of a fractal Weyl law
\cite{lu,non}. More generally, the exponent of $M$ in this law is
related to the fractal dimension of the repeller in the system.
Equation (\ref{eq:weyl}) applies under the conventional conditions
for RMT universality ($\tau_{\rm B},\lambda^{-1}\ll\tau_{\rm D}\ll
\tau_{\rm H}$), for which the low fractal dimensions can be
approximated by $d=1-1/\lambda\tau_{\rm D}+O((\lambda\tau_{\rm
D})^{-2})$ \cite{beck}. For a discussion of $d$ and the scaling
function outside this universal regime see the article of
Nonnenmacher and Zworski in the present issue \cite{non}.

\section{Quantum-to-classical crossover in the mesoscopic proximity effect \label{andreev}}

We finally consider the situation of a ballistic metallic cavity
in contact with a conventional superconductor, a so-called {\it
Andreev billiard} \cite{Kos,carlorev}. Compared to the normal
billiards considered so far, the presence of superconductivity
induces a new dynamical process called Andreev reflection, that
is, retroreflection accompanied by electron-hole conversion
\cite{Andreev}. This process prevents individual low-energy
quasiparticles from entering the superconductor.

For chaotic billiards it has been found that an excitation gap is
formed as a consequence of the Andreev reflection, in that the
Density of States (DoS) in the cavity is suppressed at the Fermi
level. The energetic scale of this gap is the ballistic Thouless
energy $E_{\rm T}=\hbar/2\tau_{\rm D}$, where $\tau_{\rm D}$ is
the average time between two consecutive Andreev reflections
\cite{Mel96}. For simplicity we consider the case of a single
superconducting terminal with $N_{\rm tot}\equiv N$ open channels
at the Fermi energy.

\subsection{Bohr-Sommerfeld quantization versus random-matrix theory}

In an ergodic cavity, all classical trajectories except a set of
zero measure eventually collide with the superconducting
interface. Andreev retroreflection is perfect at the Fermi energy,
where the hole exactly retraces the electronic path. For a nonzero
electronic excitation energy $E>0$, electron-hole symmetry is
broken, and consequently there is a mismatch between the incidence
and the retroreflection angle. For the lowest excitation energies
$\propto E_{\rm T}$ in the limit $\lambda \tau_{\rm D} \gg 1$,
this mismatch is a small parameter. Consequently,  all
electron-hole trajectories become periodic. In the semiclassical
limit where both the perimeter $L$ of the cavity and the width $W$
of the contact to the superconductor are much larger than
$\lambda_{\rm F}$ (hence $M,N\gg 1$), the semiclassical
Bohr-Sommerfeld quantization rule relates the mean DoS to the
return probability $P(t)$ to the superconductor \cite{Mel96},
\begin{equation}
\rho(E) = N \int_0^\infty {\rm d} t P(t) \sum_{m=0}^\infty
\delta\left[E-(m+\frac{1}{2}) \frac{\pi \hbar}{t}\right].
\label{bs_quant}
\end{equation}
 The shift by $1/2$ inside the $\delta$-function
 is due to two consecutive phase shifts of $\pi/2$
at each Andreev reflection. This ensures that no contributions
with an energy smaller than $E_{\rm min}(t) = \pi\hbar /2 t$
emerge from trajectories of duration $t$. Since a chaotic cavity
has an exponential distribution of return times $P(t) \propto
\exp[-t/\tau_{\rm D}]$ \cite{Bertsch}, Eq.~(\ref{bs_quant})
predicts an exponential suppression of the DoS in the chaotic case
\cite{Sch99},
\begin{equation}\label{bs_expdos}
\rho(E) = \frac{N \tau_{\rm D}}{\pi } \frac{
(2 \pi  /E \tau_{\rm D})^2 \cosh (2 \pi  /E \tau_{\rm D})}
{\sinh^2 (2 \pi  /E \tau_{\rm D})}.
\end{equation}

The DoS can also be calculated in the framework of RMT. The
excitation spectrum is obtained in the scattering approach from
the determinantal quantization condition \cite{Ben97}
\begin{equation}\label{det_andreev}
{\rm Det}[1+{\cal S}(E) {\cal S}^*(-E)] = 0.
\end{equation}
By Eq.~(\ref{heidelberg}), the scattering matrix ${\cal S}$ is
then related to a Hamiltonian matrix $H$. For low energies, Eq.\
(\ref{det_andreev}) can be transformed into an eigenvalue equation
for an effective Hamiltonian \cite{Fra96}
\begin{eqnarray}\label{andreev_eff}
{\rm Det}[E-H_{\rm eff}] = 0 \;\; , \;\;\;\;\;\;
H_{\rm eff} = \left(
\begin{array}{cc}
H & -\pi P P^T \\
 -\pi P P^T & -H^*
\end{array}
\right).
\end{eqnarray}
Assuming that $H$ is a random matrix, it has been found that the
excitation spectrum exhibits a hard gap with a ground-state energy
$E_{\rm RMT}\approx 0.6\,E_{\rm T}$ \cite{Mel96}. At first glance,
both the Bohr-Sommerfeld and the RMT approach are expected to
apply for chaotic cavities with $\lambda^{-1}$, $\tau_{\rm B}
\ll\tau_{\rm D}$. The hard gap prediction of RMT has thus to be
reconciled with the exponential suppression (\ref{bs_expdos}) from
Bohr-Sommerfeld quantization.

A path toward the solution to this {\it gap problem} was suggested
by Lodder and Nazarov \cite{lodder}, who argued that the
Bohr-Sommerfeld quantization is valid only for return times
smaller than the relevant Ehrenfest time, which was later
identified with $\tau_{\rm E}^{(2)}$ [given in Eq.\
(\ref{eq:lyapexp})] \cite{Vav03}. For $\tau_{\rm
E}^{(2)}/\tau_{\rm D} \gg 1$, it was predicted that the hard RMT
gap opens up at an energy $\simeq \hbar/\tau_{\rm E}^{(2)}$ in the
Bohr-Sommerfeld DoS. A mechanism for the opening of this gap was
soon proposed by Adagideli and Beenakker \cite{Inanc}.
Constructing a perturbation theory, they showed that diffraction
effects at the contact with the superconductor become singular in
the semiclassical limit, which results in the opening of a gap at
the inverse Ehrenfest time. More recent analytical and numerical
works confirm that the solution to the gap problem lies in the
competition between the Ehrenfest time and dwell time scales
\cite{Jac03,Vav03,Altland,Sil03b,Goo03,Goo05,kormanyos}.

\subsection{Ehrenfest suppression of the gap}

At present there are two theories for quantizing Andreev billiards
in the deep semiclassical limit. The first one proceeds along the
lines of the two-phase fluid model and the effective RMT discussed
in Section~\ref{transport}, but extended to take the energy
dependence of the scattering matrix into account
\cite{Sil03b,Goo05}. The system's scattering matrix is decomposed
into two parts,
\begin{equation}
S_0(E) = S_{\rm cl}(E) \oplus S_{\rm qm}(E),
\end{equation}
where the classical part $S_{\rm cl}(E)$ of dimension $M
(1-\exp[-\tau_{\rm E}^{(2)}/\tau_{\rm D}])$ is complemented by a quantal part
 $S_{\rm eff}(E)$ of dimension $M \exp[-\tau_{\rm E}^{(2)}/\tau_{\rm D}]$.
The excitation spectrum hence splits into
classical contributions originating from scattering trajectories
shorter than the Ehrenfest time, and quantum contributions supported
by longer trajectories for which diffraction effects are important.
An adiabatic quantization procedure allows to extract the classical
part of the excitation spectrum, while diffraction effects
are included in the theory via effective RMT,
$S_{\rm qm}(E)= \exp[i E \tau_{\rm E}^{(2)}/\hbar] S_{\rm RMT}$, where $S_{\rm RMT}$
is a random matrix from the appropriate circular
ensemble, while the factor $\exp[i E \tau_{\rm E}^{(2)}/\hbar]$ accounts 
for the delayed onset of random interference.

The second theory, due to Vavilov and Larkin \cite{Vav03}, is
based on the quasiclassical theory of  Ref.~\cite{Ale96}, which
models the mode mixing in the long time limit by isotropic residual
diffraction, with the diffraction time set to $\tau_{\rm
E}^{(2)}$. Standard techniques based on ensemble averaging can
then be applied. In the limit $\tau_{\rm E}^{(2)} \ll \tau_{\rm
D}$,the predictions of both theories for the gap value (given by
the smallest excitation energy $\epsilon_0$) differ by a factor of
$2$ \cite{carlorev},
\begin{equation}\label{gapfction}
\frac{\epsilon_0}{E_{\rm RMT}}=1-
\frac{\alpha \tau_{\rm E}^{(2)}}{2 \tau_{\rm D}}, \;\;\;\;
\alpha = \left\{
\begin{array}{cc}
2\sqrt{5}- 4& {\rm effective \; RMT}, \\
\sqrt{5}-2 & {\rm quasiclassical \; theory}.
\end{array}
\right.
\end{equation}
 In the other limit $\tau_{\rm
E}^{(2)} \gg \tau_{\rm D}$, both theories predict
$\epsilon_0=\pi\hbar/2\tau_{\rm E}^{(2)}$ \cite{carlorev}. In the
transient region $\tau_{\rm E}^{(2)} \simeq \tau_{\rm D}$, the two
theories are parametrically different. These discrepancies have
motivated detailed numerical investigations (based on the Andreev
kicked rotator described in the Appendix) which we now review 
\cite{Jac03,Goo03,Goo05}.

\begin{figure}
\vspace{0.5cm}
\begin{center}
\includegraphics[width=9cm]{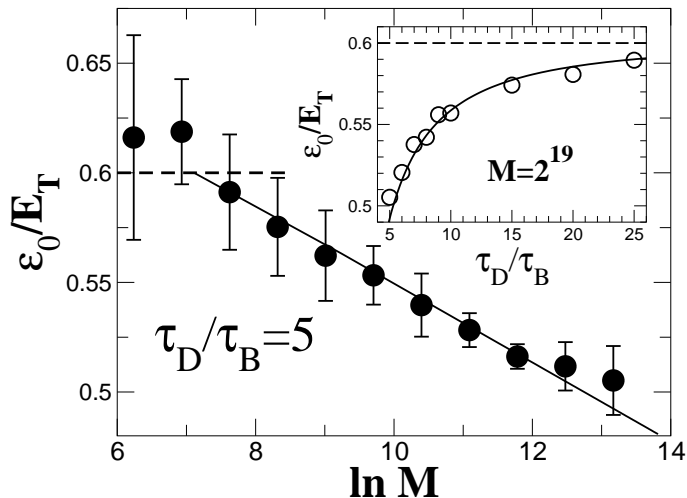}
\end{center}
\caption{\label{andreev_gap}
Main plot: Dependence of the mean Andreev gap on the system size $M$, for open kicked rotators with
$\tau_{\rm D}/\tau_{\rm B}=5$ and $K=14$. Averages have been calculated
with 400 (for $M=512$) to 40 (for $M > 5 \cdot 10^5$) different positions
of the contacts to the superconductor. The error bars represent the
root-mean-square of $\epsilon_0$. The dashed line is the
RMT prediction and the solid line is a linear fit to the data points.
Inset:
Dependence of the mean gap on $\tau_{\rm D}/\tau_{\rm B}$ for $K=14$ and
$M=524288$. The dashed line is the RMT prediction and the solid
curve is given by Eq.\ (\ref{gapfction}),
with coefficients extracted from the linear fit
in the main plot. Figure adapted from Ref.~\cite{Jac03}.}
\end{figure}

We first show in Fig.\ \ref{andreev_gap} the systematic reduction
of the excitation gap observed upon increasing the ratio
$\tau_{\rm E}^{(2)}/\tau_{\rm D}$. The data corresponds to fixed
classical configurations (dwell time and Lyapunov exponent) with
variation of the semiclassical parameter $M$. The main panel is a
semi-logarithmic plot of $\epsilon_0/E_{\rm T}$ as a function of
$M \in [2^9,2^{19}]$, for $\tau_{\rm D}/\tau_{\rm B}=5$ and
$K=14$, well in the fully chaotic regime ($\lambda\tau_{\rm B}
\approx 1.95$). The data has been fitted to
\begin{eqnarray}\label{gapfctionnum}
\frac{\epsilon_0}{E_{\rm RMT}}=1- \frac{\alpha}{2 \lambda
\tau_{\rm D}} \left[ \ln (N^2/M) - \alpha' \right],
\end{eqnarray}
as implied by Eq.~(\ref{gapfction}) (the parameter $\alpha'$
accounts for model-dependent subleading corrections to the
Ehrenfest time). We find $\alpha=0.59$ and $\alpha'=3.95$. Once
$\alpha$ and $\alpha'$ are extracted, one obtains a parameter-free
prediction for the dependence of the gap on  $\tau_{\rm
D}/\tau_{\rm B}$, which is shown as the solid line in the inset to
Fig.\ \ref{andreev_gap}. We conclude that Eq.\ (\ref{gapfction})
gives the correct parametric dependence of the Andreev gap for
small Ehrenfest times. Within the numerical uncertainties, the
value of $\alpha$ conforms with the prediction of effective RMT.
Similar conclusions were drawn in Ref.~\cite{kormanyos} from
numerical investigations of Sinai billiards.

\subsection{Quasiclassical fluctuations of the gap}

The distribution $P(\epsilon_0)$ of the Andreev gap has been calculated within
RMT in Ref.~\cite{Vav01}. It was shown to be a universal function
of the rescaled parameter $(\epsilon_0-E_{\rm RMT})/\Delta_N$,
where $\Delta_N=0.068 N^{1/3} \Delta$ gives the mean level spacing
right above the gap in terms of the bulk level spacing $\Delta$. Similarly, the standard deviation of the distribution
is given by $\sigma(\epsilon_0) = 1.27 \Delta_N$.

\begin{figure}
\begin{center}
\includegraphics[width=8cm]{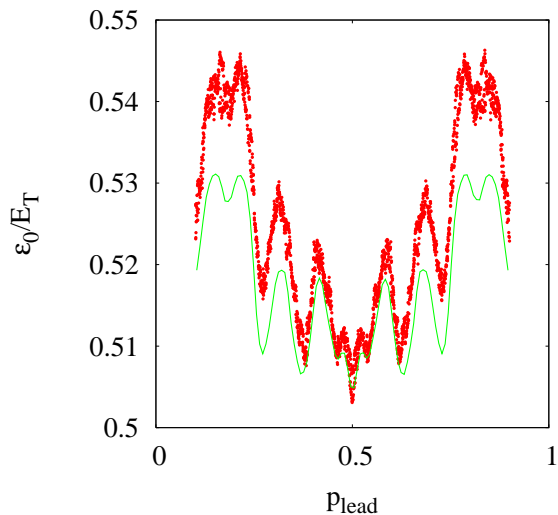}
\end{center}
\caption{\label{fig:Egap} Quantum mechanical gap values
$\epsilon_0$ of the Andreev kicked rotator as a function of the
position $p_{\rm{lead}}$ of the center of the interface with the
superconductor, for parameter values $M=131072$, $\tau_{\rm
D}/\tau_{\rm B}=5$, $K=14$. The solid line uses
effective RMT to relate the gap fluctuations to the fluctuations
of the
mean dwell time $\langle t \rangle_*$  of long classical
trajectories, defined in Eq.~(\ref{inverseT}).
Figure courtesy of Marlies Goorden \cite{Goothesis}.
}
\end{figure}

The universality of the gap distribution is violated when the
Ehrenfest time is finite. As in the case of the conductance, the
sample-to-sample gap fluctuations are then dominated by classical
fluctuations. In a simple approximation, the effective RMT model
gives a qualitative prediction for the gap value in the crossover
from a small to a large Ehrenfest time  \cite{Sil03b},
\begin{equation}\label{gap_taus}
\epsilon_0 = \frac{E_{\rm RMT}}{1+\tau_{\rm E}^{(2)}/\tau_{\rm
D}}.
\end{equation}
A more precise form of the gap function was derived in
Ref.~\cite{carlorev}. Sample-to-sample fluctuations can be
incorporated into the effective RMT model when one replaces the
dwell time in Eq.~(\ref{gap_taus}) by the mean dwell
time of long trajectories, that is one makes the substitution
\begin{equation}\label{inverseT}
(\tau_{\rm E}^{(2)} + \tau_{\rm D}) \rightarrow \langle t
\rangle_* = \int_{\tau_{\rm E}^{(2)}}^{\infty}{\rm d} t\,t P(t) \; \Bigg/ \;
\int_{\tau_{\rm E}^{(2)}}^{\infty}{\rm d} t\, P(t).
\end{equation}
This was done in
Ref.~\cite{Goo03}. The result with the correct gap function from
Ref.~\cite{carlorev} is shown in Fig.~\ref{fig:Egap}
\cite{Goothesis}. It is seen that the gap fluctuations are greatly
enhanced to the same order of magnitude as the gap itself. It was
indeed found that $\sigma(\epsilon_0)$ becomes a function of
$\tau_{\rm D}$ only in the limit of large $M$. From
Fig.~\ref{fig:Egap}, correlations between sample-to-sample
variations of $\epsilon_0$ and $\langle t \rangle_*^{-1}$ are
evident, clearly establishing the classical origin of the
sample-to-sample fluctuations in the large $M$ regime.

Together with the average value and sample-to-sample fluctuations
of $\epsilon_0$, additional numerical evidence for the validity of
the two-phase fluid model in Andreev billiards was presented in
Refs.~\cite{Goo03,Goo05}. Most notably, the critical magnetic
field at which the gap closes was found to be determined by the
competition between two values, $B_c^{\rm{eff}}$ and
$B_c^{\rm{ad}}$. These fields correspond, respectively, to the
disappearance of the gap for the quantum, effective RMT part and
the classical, adiabatically quantized part of the spectrum.
Moreover, Ref.~\cite{Goo05} showed how most of the full density of
states at finite $\tau_{\rm E}^{(2)}/\tau_{\rm D}$ can be obtained
from the effective RMT model. It would be desirable to have
similar predictions from the quasiclassical theory that could be
checked against numerical data. The excellent agreement between
the numerical data and the predictions from the effective RMT
model in Andreev billiards only adds to the intriguing controversy
about the universality of the quantal phase in the two-fluid
model, which we encountered at various places in this review.

\section{Summary and conclusions\label{conclusions}}

We gave an overview over recent theoretical and numerical
investigations which address the emergence of quantum-to-classical
correspondence in mesoscopic systems with a finite Ehrenfest time.
By now there is overwhelming evidence that the quasi-deterministic
short-time dynamics up to the Ehrenfest time jeopardizes the
universality otherwise exhibited by quantized open chaotic
systems. This was illustrated in the discussion of three different
physical situations: transport, decay of quasi-bound states, and
the mesoscopic proximity effect.

While there is consensus over the role of deterministic transport
and decay modes, there are two competing theoretical frameworks
with different predictions for the degree of wave chaos in the
long-time dynamics beyond the Ehrenfest time, namely, the
effective random-matrix theory \cite{Sil03} and the stochastic
quasiclassical theory \cite{Ale96}. Both theories incorporate the
deterministic short-time wave-packet dynamics in similar ways, and
correctly explain the suppression of shot-noise power as well as the
emergence of classical sample-to-sample fluctuations in the
semiclassical limit. However, the two theories model the long-time
dynamics in different ways, namely, via a random-matrix of reduced
dimension or via
residual diffraction. Consequently, the effective RMT predicts
that quantum interference corrections like the weak-localization
correction and parametric conductance fluctuations stay universal
deep in the semiclassical limit, while the quasi-classical theory
predicts a suppression of these effects in this limit. Moreover,
there is conflicting numerical evidence for these coherent
effects, 
with all the indications that more surprises are likely
to be uncovered. In view of this intriguing situation, more
theoretical and numerical efforts to uncover the limits of
universality in mesoscopic systems, while challenging, are clearly
desirable.

\section{Acknowledgements}

We thank \.I. Adagideli, C. Beenakker, P. Brouwer, M. Goorden, E.
Sukhorukhov, A. Tajic, J. Tworzyd\l{}o and R. Whitney for fruitful
collaborations on projects related to the topics discussed here.
C. Beenakker and P. Brouwer provided useful comments on several important points. 
M. Goorden kindly provided us with Fig.~\ref{fig:Egap} from her PhD 
thesis \cite{Goothesis}. This work has been supported by the Swiss 
National Science Foundation and the Max Planck Institute for the Physics 
of Complex Systems, Dresden.

\section*{Appendix: open kicked rotators}

The logarithmic increase of the Ehrenfest time with the effectuve
Hilbert space size $M$ requires an exponential increase in the
latter to investigate the ergodic semiclassical regime $\tau_{\rm
E} \gtrsim \tau_{\rm D}$, $\lambda \tau_{\rm D} \gg 1$, in which
deviations from RMT universality emerge due to
quantum-to-classical correspondence. The numerical results
reviewed in this paper are all obtained for a particular class of
systems, the open kicked rotator \cite{Jac03,Two03,Bor91,Oss02},
for which very efficient methods based on the fast-Fourier
transform exist. Combined with the Lanczos exact diagonalization algorithm,
as first suggested in Ref.~\cite{ketzmerick}, these methods allowed to reach
system size in excess of $M=10^6$ for the Andreev billiard 
problem \cite{Jac03}.

 The classical dynamics of the closed system are given by a symmetrized
 version of the standard
 map on the torus $x,p \in[0,2 \pi]$ \cite{Lichtenberg},
\begin{eqnarray}\label{clkrot}
\left\{\begin{array}{lll}
\bar{x} & = & x + p + \frac{K}{2} \sin x\quad(\,{\rm mod}\, 2 \pi) \\
\bar{p} & = & p + \frac{K}{2}(\sin x+ \sin\bar{x}) \quad(\,{\rm
mod}\,2 \pi).
\end{array} \right.
\end{eqnarray}
Each iteration of the map (\ref{clkrot}) corresponds to one
scattering time $\tau_{\rm B}$ off the boundaries of a fictitious
cavity. The dynamics of this system ranges from fully integrable
($K=0$) to well-developed chaos [$K \ge 7$, with Lyapunov exponent
$\lambda\approx\tau_{\rm B}^{-1}\ln (K/2)$].

The map (\ref{clkrot}) can be quantized by discretization of the
space coordinates $x_m=2 \pi m/M$, $m=1,\ldots M$. The quantum
representation is then provided by a unitary $M \times M$ Floquet
operator $F$ \cite{Haake}, which gives the time evolution for one
iteration of the map. For our specific choice of the kicked
rotator, the Floquet operator has matrix elements
\begin{eqnarray}\label{kickedU}
F_{m,m'} &=& M^{-1/2} \exp \{-(iMK/4\pi) [\cos(2\pi m/M)+\cos(2\pi m'/M)] \}
\nonumber \\
& & \times \sum_l \exp[2 \pi i l(m-m')/M] \exp[-(\pi i/2M) l^{2}].
\end{eqnarray}
The spectrum $\exp(-i E_n\tau_B/\hbar)$ of $F$ defines a discrete
set of $M$ quasienergies $E_n \in[0,h/\tau_B)$ with an average
level spacing $\Delta = h/M\tau_B$.

For the transport problem, the system is opened up by
defining two ballistic openings via absorbing phase space strips
$[x_L-\delta x,x_L+\delta x]$ and $[x_R-\delta x,x_R+\delta x]$,
each of a width $2 \delta x=\pi\tau_{\rm B}/\tau_{\rm D}$.
Much in the same way as in the Hamiltonian case \cite{Guh98}, a
quasienergy-dependent $2N \times 2N$ scattering matrix can be determined
from the Floquet operator $F$ as \cite{fyodorov}
\begin{equation}\label{smatrix}
{\cal S}(E) = P [\exp(-i E\tau_B/\hbar) - F (1-P^T P)]^{-1} F P^T.
\end{equation}
The $2N \times M$-dimensional matrix $P$
describes the coupling to the leads, and is given by
\begin{eqnarray}\label{lead}
 P_{n,m}=\left\{\begin{array}{ll}
1& \mbox{if $n=m \in \{(x_{\rm R,L}-\delta)M/2\pi,x_{\rm R,L}+\delta)M/2\pi\}$}\\
0& \mbox{otherwise.}
\end{array}\right.
\end{eqnarray}
The number of channels in each opening is given by $N=\,{\rm
Int}\,[\delta M/\pi]$. An ensemble of samples with the same
microscopic properties can be defined by varying the position
 of the two openings for fixed
$\tau_{\rm D}/\tau_{\rm B}$ and $K$, or by varying the energy $E$.

For the escape problem, $P$ couples only to a single opening, and
the quasibound states are obtained by diagonalization of the
truncated quantum map $(1-P^TP) F$.

So far we described particle excitations in a normal metal. In
order to model an Andreev billiard \cite{Jac03}, we also need hole
excitations. A particle excitation with energy $E_{m}$ (measured
relatively to the Fermi level) is identical to a hole excitation
with energy $-E_{m}$ which propagates backwards in time.  This
means that hole excitations in a normal metal have Floquet
operator $F^{\ast}$. Andreev reflections occurs at the opening to
the superconducting reservoir, which is again represented by the
matrix $P$. The symmetrized quantum Andreev map is finally
constructed from the matrix product
\begin{eqnarray}
{\cal F}={\cal P}^{1/2}
\left(\begin{array}{cc}
F&0\\
0&F^{\ast}
\end{array}\right) {\cal P}^{1/2} ,\;\;
{\cal P}=\left(\begin{array}{cc}
1-P^{\rm T}P&-iP^{\rm T}P\\
-iP^{\rm T}P&1-P^{\rm T}P
\end{array}\right).\label{calFdef}
\end{eqnarray}
The excitation spectrum is obtained by diagonalization of ${\cal
F}$, whose quasienergy spectrum exhibits two gaps at $E = 0$ and
$E = h/2\tau_{\rm B}$. It can be shown that the excitation
spectrum is identical to the solutions of the conventional
determinantal equation ${\rm det}\,[1+{\cal S}(E){\cal
S}^*(E)]=0$, where the scattering matrix is given by Eq.\
(\ref{smatrix}).
\\[2cm]

\end{document}